# Zigzag-base Folded Sheet Cellular Mechanical Metamaterials

Maryam Eidini

**Affiliations**

Department of Civil and Environmental Engineering, University of Illinois at Urbana Champaign, 205 North Mathews Ave., Urbana, IL 61801, USA.

Correspondence to:  eidinin1@illinois.edu

**Abstract**

The Japanese art of turning flat sheets into 3D intricate structures, origami, has inspired design of mechanical metamaterials. Mechanical metamaterials are artificially engineered materials with uncommon properties. Miura-ori is a remarkable origami folding pattern with metamaterial properties and a wide range of applications. In this study, by dislocating the zigzag strips of a Miura-ori pattern along the joining ridges, we create a class of one-degree of freedom (DOF) cellular mechanical metamaterials. The resulting configurations are based on a unit cell in which two zigzag strips surround a hole with a parallelogram cross section. We show that dislocating zigzag strips of the Miura-ori along the joining ridges, preserves and/or tunes the outstanding properties of the Miura-ori. The introduced materials are lighter than their corresponding Miura-ori sheets due to the presence of holes in the patterns. Moreover, they are amenable to similar modifications available for Miura-ori which make them appropriate for a wide range of applications.

**Introduction**

Miura-ori, a zigzag/herringbone-base origami folding pattern, has attracted substantial attention in science and engineering for its remarkable properties [1, 2, 3, 4, 5]. The exceptional mechanical properties of the Miura-ori [4, 6], the ability to produce its morphology as a self-organized buckling pattern [1, 2] and its geometric adaptability [7, 8] has made the pattern suited for applications spanning from metamaterials [4] to fold-core sandwich panels [9]. Moreover, Miura-ori is a mechanical metamaterial with negative Poisson's ratio for a wide range of its geometric parameters [10, 11]. Mechanical metamaterials are artificially engineered materials with unusual material properties arising from their geometry and structural layout. Poisson's ratio is defined as the negative ratio of transverse to axial strains. Poisson's ratios of many common isotropic elastic materials are positive, i.e., they expand transversely when compressed in a given direction. Conversely, when compressed, materials with negative Poisson's ratio or auxetics contract in the directions perpendicular to the applied load. Discovery and creating of auxetic materials has been of interest due to improving the material properties of auxetics [12, 13, 14, 15]. Auxetic behavior may be exploited through rotating rigid and semi-rigid units [16, 17], chiral structures [18, 19], reentrant structures [20, 21, 22], elastic instabilities in switchable auxetics [23, 24], creating cuts in materials [25], and in folded sheet materials [4, 10]. The latter is the concentration of the current research.

Research studies have shown that the herringbone geometry leads to auxetic properties in folded

---

sheet materials [4, 10] and textiles [26, 27], and its morphology arises in biological systems [28, 29, 30]. Due to possessing unprecedented deformability, the herringbone structure fabricated by bi-axial compression, has been also used in deformable batteries and electronics [31, 5, 32]. Kirigami, the art of paper cutting, has been applied as 3D core cellular structures and solar cells among others [33, 34]. The current research expands on a recent study by Eidini and Paulino [10] where origami folding has been combined with cutting patterns to create a class of cellular metamaterials. In the present study, we use the concept of the Poisson's ratio of a one-DOF zigzag strip (i.e., $v_z = -\tan^2\phi$) described by Eidini and Paulino [10] which provides inspiration to tune and/or preserve the properties of the Miura-ori. In this regard, by dislocating the zigzag strips of the Miura-ori pattern along the joining fold lines, we create a novel class of metamaterials. The resulting configurations are based on a one-DOF unit cell in which two zigzag strips surround a hole with a parallelogram cross section.

## Geometry of the Patterns

As shown in Fig. 1, arrangement of the zigzag strips with offsets creates the parallelogram holes in the patterns.

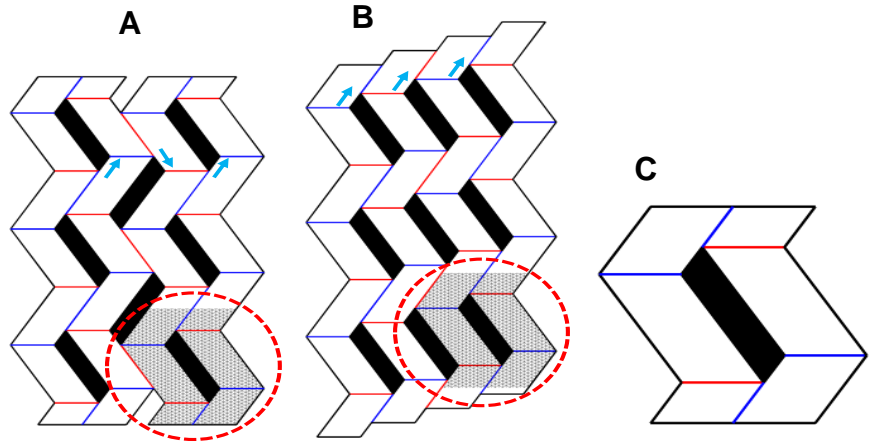

**Fig. 1. Crease patterns of sample zigzag-base folded materials introduced in the current work and their unit cell.** (**A**) Changing the direction of the offset from a zigzag strip to the next adjoining one results in a pattern with the holes located in different directions - the direction of the offsets are shown with blue arrows. (**B**) Arranging the offsets all to one side, results in zigzag strips with the holes all located with the same direction. (**C**) Crease pattern of the unit cell. In the figures, the blue and red lines show mountain and valley folds, respectively, and hatched black areas represent the places of the cuts.

The *Zigzag unit Cell with Hole* (ZCH) of the patterns is shown in Fig. 1C and is parametrized in Fig. 2A. The equations defining the geometry of the ZCH are given by

$$w = 2b\sin\phi \qquad \ell = 2a\frac{\cos\alpha}{\cos\phi} \qquad h = a\sin\alpha\sin\theta \qquad b_0 = (b - b_h)/2 \qquad (1)$$

The expression relating the angle $\phi$ and the fold angle $\theta$ is as follows

$$\tan\phi = \cos\theta \tan\alpha \qquad (2)$$

If considered in the context of rigid origami, the ZCH is a one-DOF mechanism system. Sample patterns containing ZCH unit cells are presented in Fig. 3, and the patterns have one DOF. We obtain the DOF of the patterns in this work using the approach mentioned in [10].

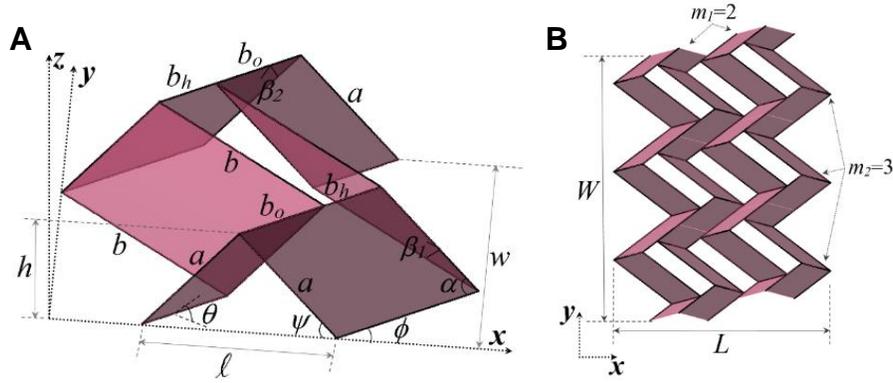

**Fig. 2. Geometry of ZCH pattern.** (**A**) Geometry of the unit cell. The geometry of a ZCH sheet can be parameterized by the geometry of a parallelogram facet, hole width $b_h$, and fold angle $\phi \in [0, \alpha]$ which is the angle between the edges $b_0$ (and $b$) and the $x$-axis in the $xy$-plane. Other important angles in the figure are fold angle between the facets and the $xy$-plane, i.e., $\theta \in [0, \pi/2]$; angle between the fold lines $a$ and the $x$-axis, i.e., $\psi \in [0, \alpha]$; Dihedral fold angles between parallelogram facets $\beta_1 \in [0, \pi]$ and $\beta_2 \in [0, \pi]$, joining along fold lines $a$ and $b_0$, respectively. (**B**) A ZCH sheet with $m_1=2$ and $m_2=3$ and outer dimensions $L$ and $W$.

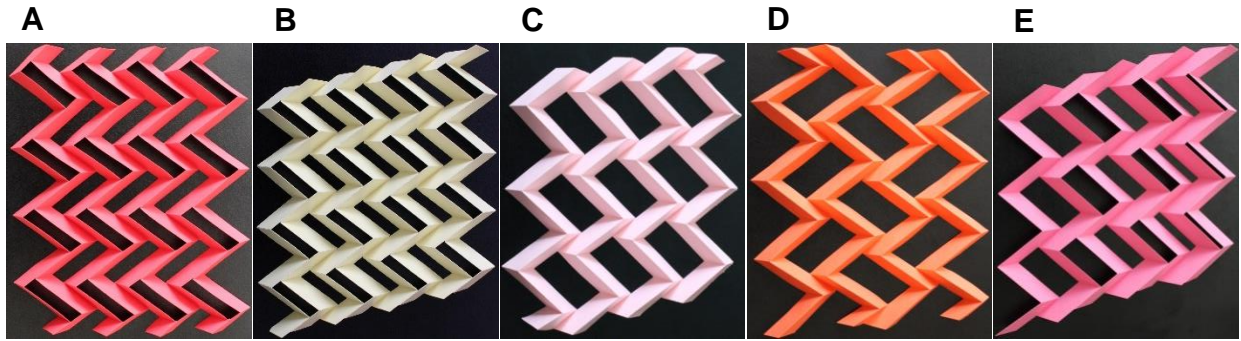

**Fig. 3. Sample patterns of ZCH.** (**A**-**E**) Sample ZCH sheets created by changing the direction and/or the amount of the offsets in the patterns. Note that by changing the height $h$, the length and width of the parallelogram facets, the hole width (pattern E) and other changes (e.g., similarly to the Miura-ori, changing the geometry of the facets to get the curved version and others) we can produce numerous graded and/or shape morphing materials/structures.

## Key Mechanical Behaviors of the Patterns

Being in the class of zigzag-base patterns with one-DOF planar mechanism, the patterns of ZCH

shown in Fig. 3 all have $\upsilon_z$ equal to $-tan^2\phi$ [10]. Using the outer dimensions of the sheet, the Poisson's ratio of a regular ZCH sheet (for example, sample patterns shown in Fig. 2B and Fig. 3A) is given by

$$\left(\upsilon_{WL}\right)_{e-e} = -\frac{\varepsilon_L}{\varepsilon_W} = -\frac{dL/L}{dW/W} = -\tan^2\phi \frac{\eta\cos\alpha - \cos^2\phi}{\eta\cos\alpha + \cos^2\phi} \quad (3)$$

in which

$$\eta = \frac{m_1 a}{m_1 b_h + b_0} \quad (4)$$

Hence, the Poisson's ratio of a repeating unit cell (in an infinite tessellation) is given by

$$\left(\upsilon_\infty\right)_{e-e} = \left(\upsilon_{L_r w}\right)_{repeating\,unit\,cell} = -\frac{\varepsilon_{L_r}}{\varepsilon_w} = -\frac{dL_r/L_r}{dw/w} = -\tan^2\phi \frac{a\cos\alpha - b_h \cos^2\phi}{a\cos\alpha + b_h \cos^2\phi} \quad (5)$$

The value presents the Poisson's ratio of a regular ZCH sheet for an infinite configuration as well. The value is positive if $a\cos\alpha < b_h \cos^2\phi$. The value is negative if $a\cos\alpha > b_h \cos^2\phi$. If $a/b_h \to \infty$, the Poisson's ratio of a repeating unit cell approaches $\upsilon_z$. If $b/a \to \infty$, the Poisson's ratio of a repeating unit cell of the Miura-ori remains as $-tan^2\phi$, i.e., the $\upsilon_z$ (Fig. 4B), but the Poisson's ratio of a repeating unit cell of ZCH approaches $tan^2\phi$. This phenomenon happens due to the existence of the holes in the ZCH patterns.

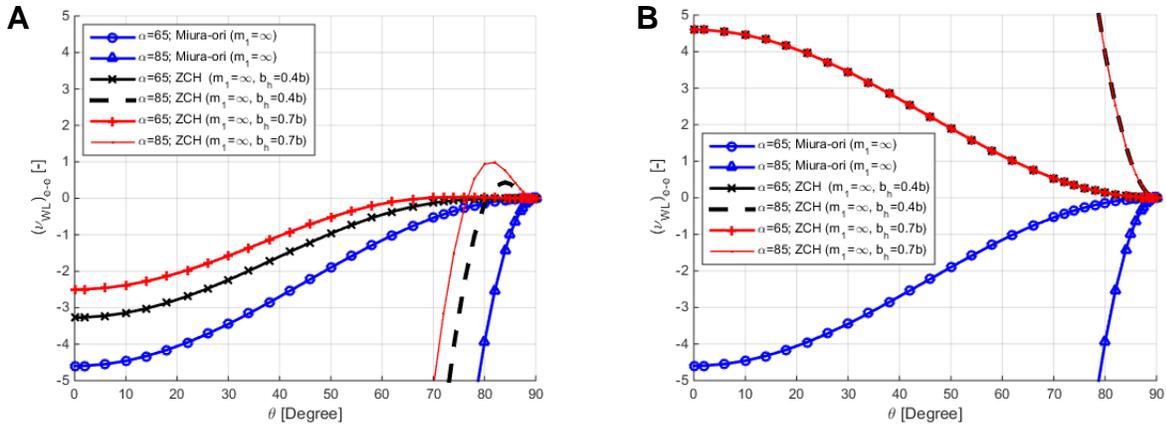

**Fig. 4. In-plane Poisson's ratio of metamaterials introduced in this work with infinite configurations**. (**A**) Poisson's ratio of Miura-ori and ZCH sheets for ($m_1 = \infty$) and $a=b$, and two different hole widths - the values correspond to the Poisson's ratios of the repeating unit cells of sheets as well. (**B**) Poisson's ratio of repeating unit cells of ZCH and Miura-ori sheets if $b/a \to \infty$.

Upon bending, a ZCH sheet exhibits a saddle-shaped deformation (see Fig. A1) which is a property for materials with positive Poisson's ratio [15]. Using the bar-framework numerical approach [4], the results of eigen-value analyses of sample ZCH patterns reveal similar behavior to those observed in Miura-ori and $BCH_n$ [4, 10] (see Fig. 5 and Appendix for more details).

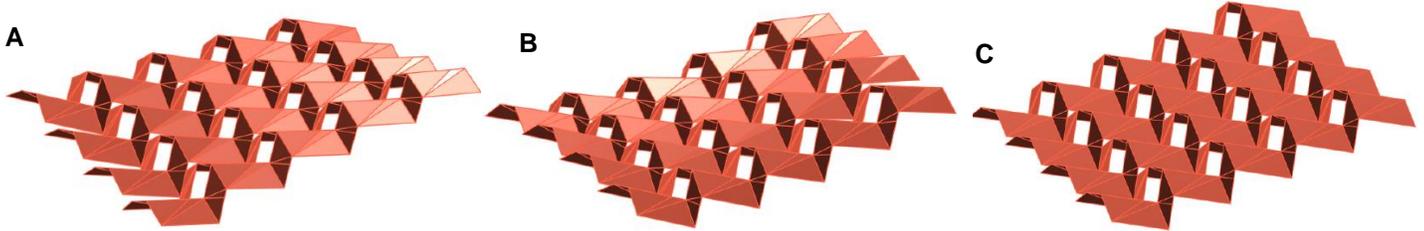

**Fig. 5. Behavior of ZCH sheets under bending and the results of eigenvalue analyses of sample ZCH sheets.** (**A**) Twisting, (**B**) saddle-shaped and (**C**) rigid origami behavior (planar mechanism) of a 4x4 and a 4x3 patterns of ZCH with the holes located on various directions ($a$=1; $b$=2; $\alpha$=60°).

## Concluding Remarks

In this study, we have presented a method to tune and/or keep the properties of the Miura-ori, i.e., the most remarkable origami folding pattern. The resulting configurations are zigzag-base patterns which are flat-foldable, developable and 1-DOF systems. The main advantages of the patterns are highlighted as follows: (i) the patterns are amenable to similar modifications and/or applications available for the Miura-ori (e.g., [4, 35, 8, 36, 37] - see Fig. A2 to A4). (ii) Due to possessing holes in their configurations, they are less dense than their corresponding Miura-ori patterns (see Fig. A5 to Fig. A7). (iii) They extend geometrical and mechanical design space of the prior known zigzag-base patterns such as Miura-ori and $BCH_2$ (for example, see Fig. 4). (iv) Despite existence of the holes, they are all single-degree of freedom (SDOF) systems for the rigid origami behavior. SDOF rigid mechanisms are appropriate for low energy, efficient and controllable deployable structures. (v) Compared with $BCH_n$ patterns [10] whose unit cell includes two large and $2n$ small parallelogram facets, the unit cell of the patterns introduced in this work possess identical number of facets on each side of the hole. Hence, they can be more appropriate than their corresponding $BCH_n$ patterns when the base material of the facets is thicker (e.g., when separate thick rigid panels are connected with frictionless hinges). (vi) For applications such as sandwich folded-cores, unlike *Zeta core* [38], the patterns remain developable by adding surfaces at the top and bottom of the patterns to increase the bonding areas (see Fig. A2) - developable sheets are well-suited for continuous manufacturing techniques available for folded core structures.

In summary, the characteristics of the introduced patterns make them suitable for a broad range of applications from folded-core sandwich panels and morphing structures to metamaterials at various length scales.

# References:

[1] L. Mahadevan and S. Rica, "Self-organized origami," *Science,* vol. 307, p. 1740, 2005.

[2] K. Miura, "The science of miura-ori," in *4th International Meeting of Origami Science, Mathematics, and Education, RJ Lang, Ed., AK Peters*, 2009.

[3] K. Tanizawa and K. Miura, "Large displacement configurations of bi-axially compressed infinite plate," *Japan Society for Aeronautical and Space Sciences, Transactions,* vol. 20, pp. 177-187, 1978.

[4] M. Schenk and S. D. Guest, "Geometry of miura-folded metamaterials," *Proceedings of the National Academy of Sciences,* vol. 110, no. 9, pp. 3276-3281, 2013.

[5] Z. T. M. Song, R. Tang, Q. Cheng, X. Wang, D. Krishnaraju, R. Panat, C. K. Chan, H. Yu and H. Jiang, "Origami lithium-ion batteries," *Nature communications,* vol. 5, p. 3140 , 2014.

[6] Z. Y. Wei, Z. V. Guo, L. Dudte, H. Y. Liang and L. Mahadevan, "Geometric mechanics of periodic pleated origami," *Physical Review Letters,* vol. 110, no. 21, pp. 215501-215505, 2013.

[7] T. Tachi, "Generalization of rigid foldable quadrilateral mesh origami," in *Proc. of the Int. Association for Shell and Spatial Structures (IASS) Symp*, Valencia, Spain, 2009.

[8] F. Gioia, D. Dureisseix, R. Motro and B. Maurin, "Design and analysis of a foldable/unfoldable corrugated architectural curved envelop," *Journal of Mechanical Design,* vol. 134, no. 3, p. 031003, 2012.

[9] S. Heimbs, "Foldcore sandwich structures and their impact behaviour: An overview," in *Dynamic Failure of Composite and Sandwich Structures*, Springer Netherlands, 2013, pp. 491-544.

[10] M. Eidini and G. H. Paulino, "Unravelling Metamaterial Properties in Zigzag-base Folded Sheets," *Science Advances,* vol. 1, no. 8, p. e1500224, 2015.

[11] C. Lv, D. Krishnaraju, G. Konjevod, H. Yu and H. Jiang, "Origami based Mechanical Metamaterials," *Scientific Reports,* vol. 4, p. 5979, 2014.

[12] R. Lakes, "Foam structures with a negative Poisson's ratio," *Science,* vol. 235, no. 4792, pp. 1038-1040, 1987.

[13] K. E. Evans and A. Alderson, "Auxetic materials: functional materials and structures from lateral thinking!," *Advanced materials ,* vol. 12, no. 9, pp. 617-628, 2000.

[14] W. Yang, Z.-M. Li, W. Shi, B.-H. Xie and M.-B. Yang, "Review on auxetic materials," *Journal of materials science,* vol. 39, no. 10, pp. 3269-3279, 2004.

[15] A. Alderson and K. L. Alderson, "Auxetic Materials," *Proceedings of the Institution of Mechanical Engineers, Part G: Journal of Aerospace Engineering,* vol. 221, no. 4, pp. 565-575, 2007.

[16] J. N. Grima, A. Alderson and K. E. Evans, "Auxetic behaviour from rotating rigid units," *Physica status solidi (b),* vol. 242, no. 3, pp. 561-575, 2005.

[17] J. N. Grima, V. Zammit, R. Gatt, A. Alderson and K. E. Evans, "Auxetic behaviour from rotating semi-rigid units," *Physica Status Solidi (b) ,* vol. 244, no. 3, pp. 866-882, 2007.

[18] D. Prall and R. S. Lakes, "Properties of a chiral honeycomb with a Poisson's ratio of -1," *International Journal of Mechanical Sciences,* vol. 39, no. 3, pp. 305-314, 1997.

[19] A. Spadoni and M. Ruzzene, "Elasto-static micropolar behavior of a chiral auxetic lattice," *Journal of the Mechanics and Physics of Solids,* vol. 60, no. 1, pp. 156-171, 2012.

[20] R. F. Almgren, "An isotropic three-dimensional structure with Poisson's ratio -1," *Journal of Elasticity ,* vol. 15, no. 4, pp. 427-430, 1985.

[21] K. E. Evans, A. Alderson and F. R. Christian, "Auxetic two-dimensional polymer networks. An example of tailoring geometry for specific mechanical properties," *J. Chem. Soc., Faraday Trans.,* vol. 91, no. 16, pp. 2671-2680, 1995.

[22] H. Yasuda and J. Yang, "Reentrant Origami-Based Metamaterials with Negative Poisson's Ratio and Bistability," *Physical Review Letters,* vol. 114, no. 18, p. 185502, 2015.

[23] K. Bertoldi, P. M. Reis, S. Willshaw and T. Mullin, "Negative Poisson's ratio behavior induced by an elastic instability," *Advanced Materials,* vol. 22, no. 3, pp. 361-366, 2010.

[24] S. Babaee, J. Shim, J. C. Weaver, E. R. Chen, N. Patel and K. Bertoldi, "3D soft metamaterials with negative Poisson's ratio," *Advanced Materials,* vol. 25, no. 36, pp. 5044-5049, 2013.

[25] Y. Cho, J.-H. Shin, A. Costa, T. A. Kim, V. Kunin, J. Li, S. Y. Lee, S. Yange, H. Nam Hanf, I.-S. Choi and D. J. Srolovitze, "Engineering the shape and structure of materials by fractal cut," *Proceedings of the National Academy of Sciences ,* vol. 111, no. 49, pp. 17390-17395, 2014.

[26] Z. Wang and H. Hu, "3D auxetic warp-knitted spacer fabrics," *physica status solidi (b),* vol. 251, no. 2, pp. 281-288, 2014.

[27] M. Glazzard and P. Breedon, "Weft-knitted auxetic textile design," *Physica Status Solidi (b),* vol. 251, no. 2, pp. 267-272, 2014.

[28] M. B. Amar and F. Jia, "Anisotropic growth shapes intestinal tissues during embryogenesis," *Proceedings of the National Academy of Sciences,* vol. 110, no. 26, pp. 10525-10530, 2013.

[29] A. E. Shyer, T. Tallinen, N. L. Nerurkar, Z. Wei, E. S. Gil, D. L. Kaplan, C. J. Tabin and L. Mahadevan, "Villification: how the gut gets its villi," *Science,* vol. 342, no. 6155, pp. 212-218, 2013.

[30] H. Kobayashi, B. Kresling and J. F. V. Vincent, "The geometry of unfolding tree leaves," *Proceedings of the Royal Society of London. Series B: Biological Sciences,* vol. 265, pp. 147-154, 1998.

[31] J. A. Rogers, T. Someya and Y. Huang, "Materials and mechanics for stretchable electronics," *Science,* vol. 327, no. 5973, pp. 1603-1607, 2010.

[32] Y. Zhang, Y. Huang and J. A. Rogers, "Mechanics of stretchable batteries and supercapacitors," *Current Opinion in Solid State and Materials Science,* 2015.

[33] T. Nojima and K. Saito, "Development of newly designed ultra-light core structures," *JSME International Journal Series A,* vol. 49, no. 1, pp. 38-42, 2006.

[34] A. Lamoureux, K. Lee, M. Shlian, S. R. Forrest and M. Shtein, "Dynamic kirigami structures for integrated solar tracking," *Nature communications,* p. 6:8092, 2015.

[35] K. C. Cheung, T. Tachi, S. Calisch and K. Miura, "Origami interleaved tube cellular materials," *Smart Materials and Structures ,* vol. 23, no. 9, p. 094012, 2014.

[36] J. M. Gattas, W. Wu and Z. You, "Miura-Base Rigid Origami: Parameterizations of First-Level Derivative and Piecewise Geometries," *Journal of Mechanical Design,* vol. 135, no. 11, p. 111011, 2013.

[37] M. Schenk, S. D. Guest and G. J. McShane, "Novel stacked folded cores for blast-resistant sandwich beams," *International Journal of Solids and Structures,* vol. 51, no. 25, pp. 4196-4214, 2014.

[38] K. Miura, "Zeta-Core Sandwich- Its Concept and Realization," *Inst. of Space and Aeronautical Science, University of Tokyo,* vol. 480, pp. 137-164, 1972.

# APPENDIX

## 1- Out-of-plane behavior of the patterns

Upon bending, ZCH sheets exhibit saddle-shaped curvatures (Fig. A1). The behavior is typically observed in materials with positive Poisson's ratio [15].

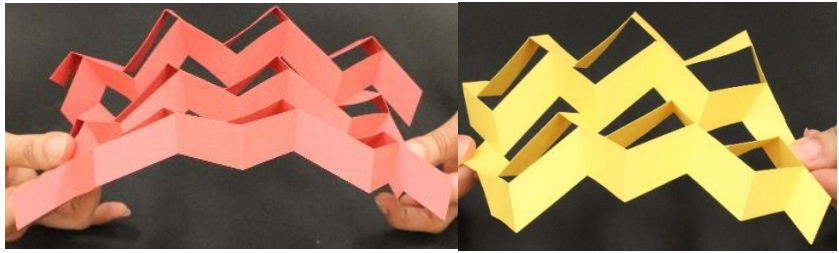

**Fig. A1.** Behavior of sheets of the ZCH patterns under bending. Sheets of ZCH deform into saddle-shaped curvature under bending.

## 2- Other variations of ZCH and their assemblages

The variation of the ZCH pattern shown in Fig. A2 provides additional bonding areas on the crests of the corrugation for applications as folded-core sandwich panels. Furthermore, by changing the geometry of the facets similarly to the curved version of the Miura-ori [36], we can create a curved version shown in Fig. A3. In addition to the variations available for the Miura-ori (e.g., [7, 36]) which are applicable to these patterns, changing the hole width $b_h$ (Fig. 3F) and the width of the bonding areas at the crests (shown in Fig. A2) combined with other changes (e.g., the variations mentioned in references [7, 36]) can provide extensive flexibility to create various variations/shapes from the patterns.

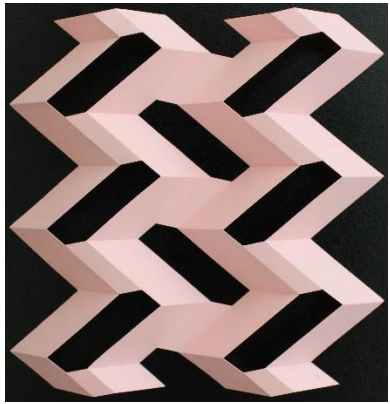

**Fig. A2.** A developable ZCH pattern with augmented bonding areas.

    

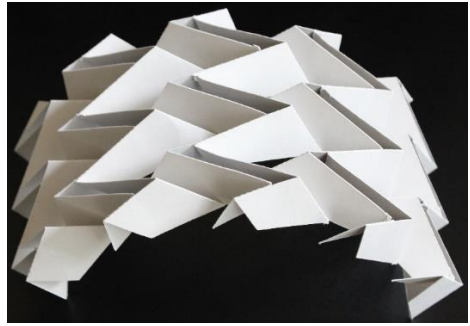

**Fig. A3.** A curved ZCH pattern.

Sample one-DOF cellular materials designed based on the ZCH patterns are shown in Fig. A4. The stacked materials shown in Fig. A4 A and B are appropriate for applications such as impact absorbing devices [37]. The interleaved ZCH tubular material shown in Fig. A4C is similar to the one made from the Miura-ori pattern [35], which is a bi-directionally flat-foldable material. Its geometry results in a material which is soft in two directions and relatively stiff in the third direction. The samples shown in Fig. A4D to Fig. A4F are bi-directionally flat-foldable materials made from different assemblages of the ZCH tubes.

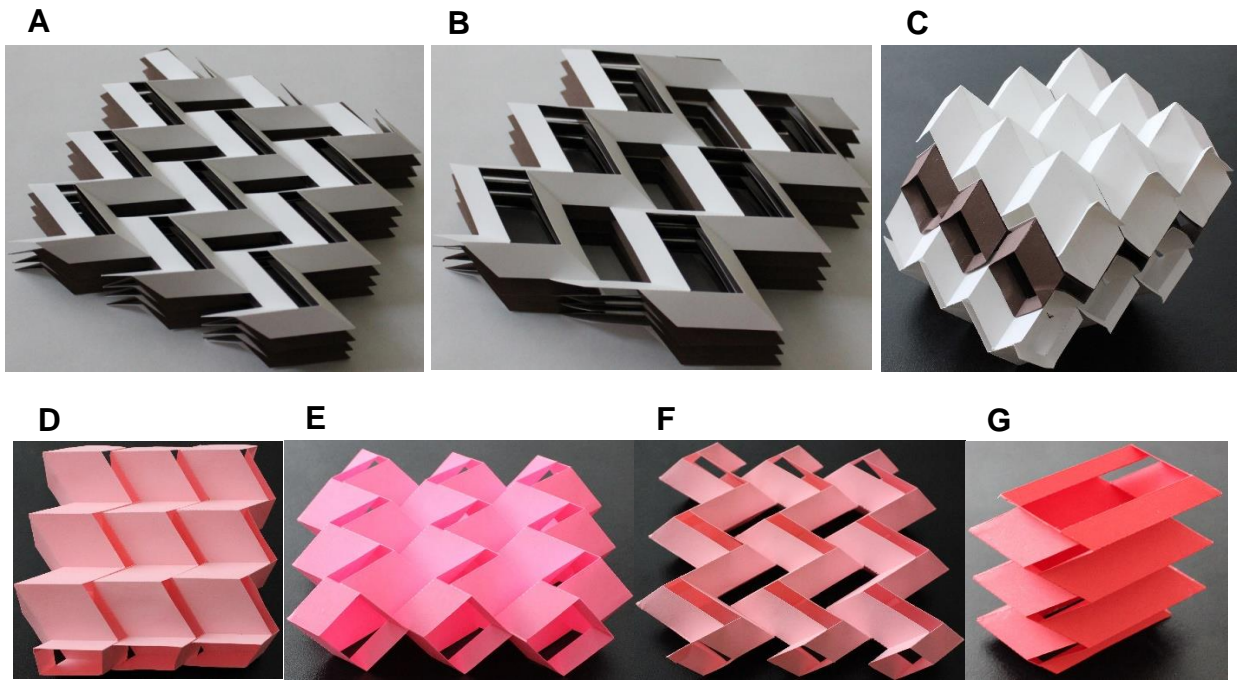

**Fig. A4. Cellular foldable metamaterials.** (**A, B**) Stacked cellular metamaterials made from 7 layers of folded ZCH sheets. Each material includes two different geometries of similar sheets. (**C**) Interleaved ZCH tubular materials. (**D-F**) Materials made from various assemblages of ZCH tubes. (**G**) Sample ZCH tube with a rectangular cross section.

 

## 3- Comparison of ZCH with Miura-ori and BCH$_2$ for the same amount of mass

In this section, we compare the density of the ZCH with its corresponding Miura-ori and BCH$_2$ sheets (see Fig. A5 for sample sheets). Knowing that widths and heights of the corresponding sheets are identical, the ratio of the density of the Miura-ori to that of the ZCH is given by

$$\rho_{Miura/ZCH} = \frac{V_{ZCH}}{V_{Miura}} = \frac{L_{ZCH}}{L_{Miura}} \quad (1)$$

Equation (4) in the main text gives the length of the ZCH sheet. Note that considering $b_h = 0$ in that relation, gives the length of the Miura-ori. The ratio is shown in Fig. A6. Similarly, for BCH$_2$, the ratio of the density of the BCH$_2$ to that of the ZCH is shown in Fig. A7. The length of the hole, *i.e.* $b_h$, at which the density of the ZCH is equal to that of the BCH$_2$ is given in the following.

$$b_h = \frac{b(m_1 - 1)}{2(2m_1 - 1)} \quad (2)$$

For the infinite configuration of the sheets (*i.e.*, $m_1 \to \infty$), the length of the hole at which the density of ZCH is equivalent to that of the BCH$_2$ is given by

$$b_h = \frac{1}{4}b \quad (3)$$

Consequently, for $b_h > 0.25b$, the ZCH is less dense than its corresponding BCH$_2$ sheet and for $b_h < 0.25b$, BCH$_2$ is less dense than its corresponding ZCH sheet.

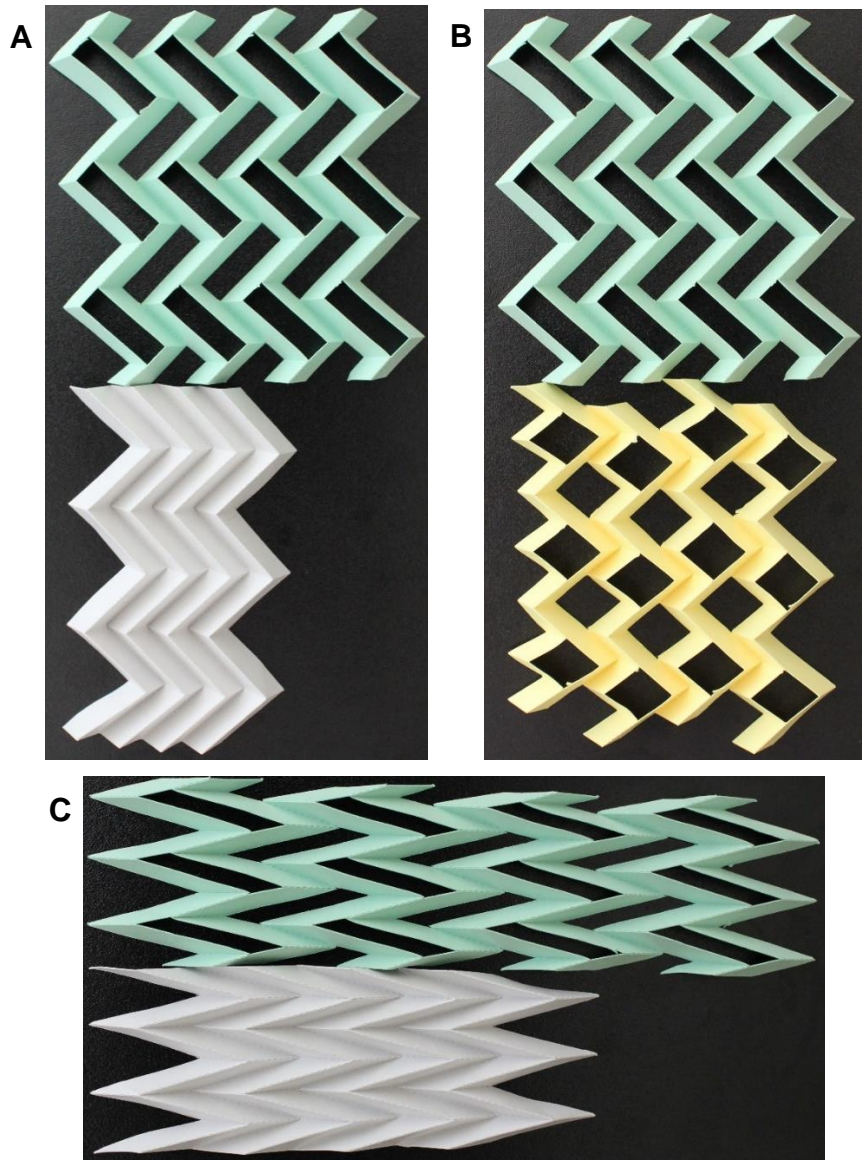

**Fig. A5.** Sample ZCH with its corresponding Miura-ori and BCH$_2$ sheets. (**A**) A 4x3 ZCH sheet with its corresponding Miura-ori with $b=2a$, $\alpha=70°$ and $b_h=0.3b$. (**B**) A 4x3 ZCH sheet with its corresponding BCH$_2$ sheet with $b=2a$, $\alpha=70°$ and $b_h=0.3b$. (**C**) A 4x3 ZCH sheet with its corresponding Miura-ori with $b=2a$, $\alpha=30°$ and $b_h=0.3b$.

         

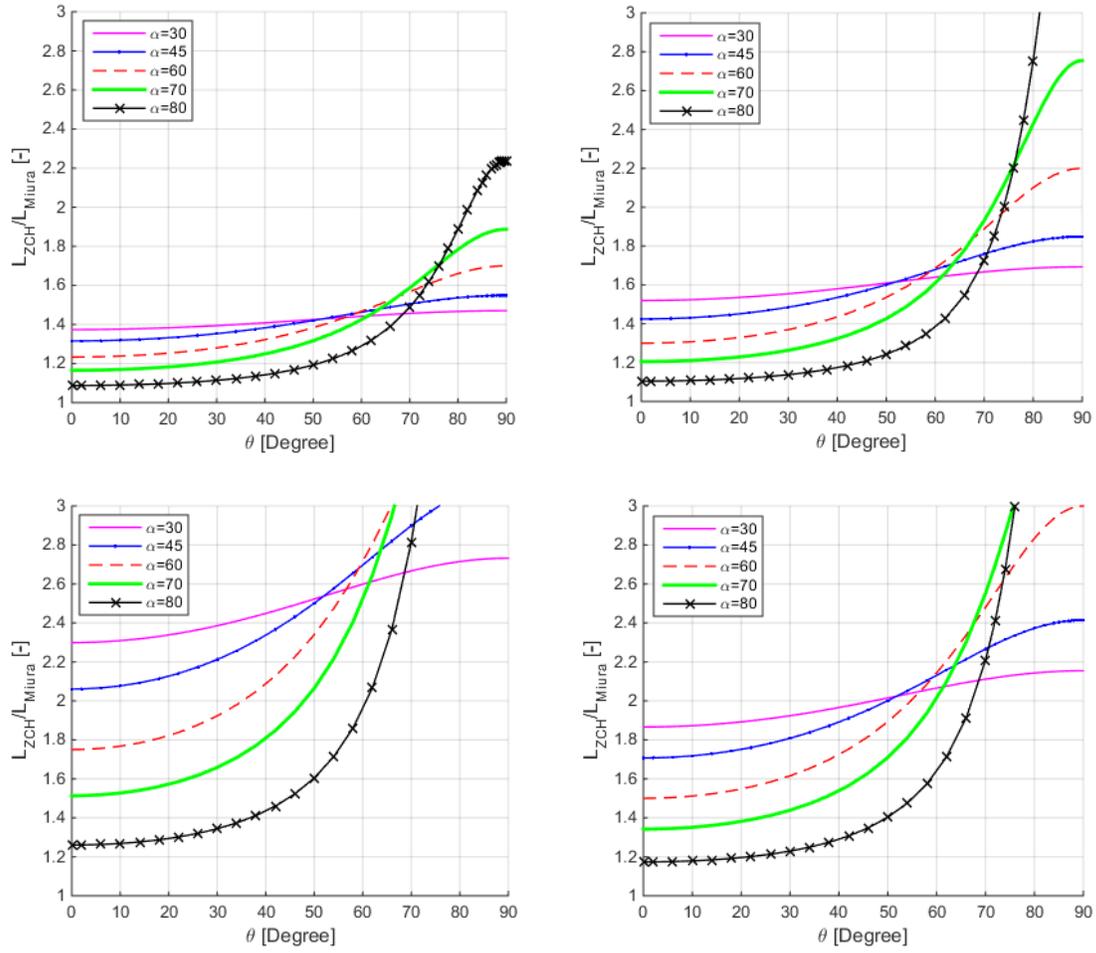

**Fig. A6.** Density of ZCH compared with that of its corresponding Miura-ori. (**A**) $m_1=4$, $b/a=2$ and $b_h=0.3b$. (**B**) $m_1=\infty$, $b/a=2$ and $b_h=0.3b$. (**C**) $m_1=\infty$, $b/a=5$ and $b_h=0.3b$. (**D**) $m_1=\infty$, b/a=2 and $b_h=0.5b$.

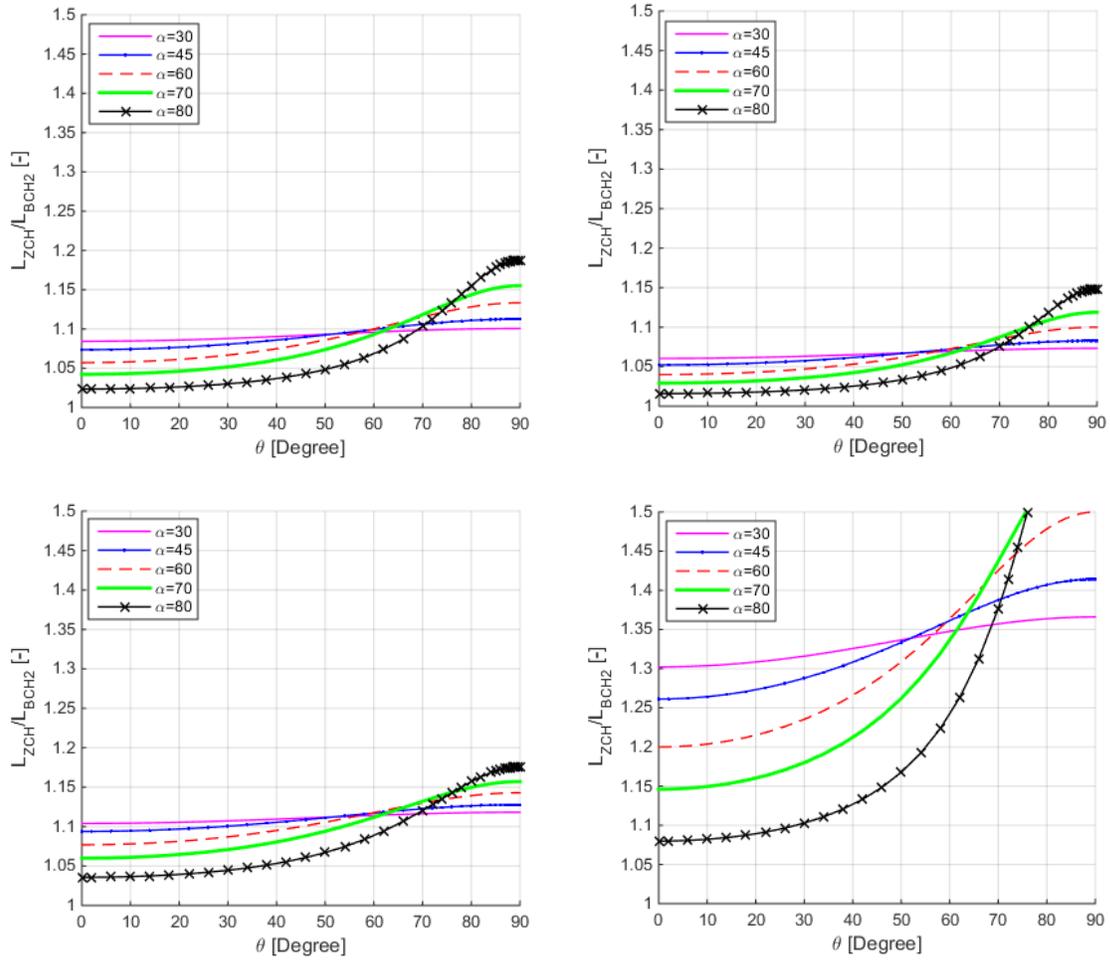

**Fig. A7.** Density of ZCH compared with that of its corresponding $BCH_2$ sheet. (**A**) $m_1$=4, $b/a$=2 and $b_h$=0.3$b$. (**B**) $m_1$=∞, $b/a$=2 and $b_h$=0.3b. (**C**) $m_1$=∞, $b/a$=5 and $b_h$=0.3$b$. (**D**) $m_1$=∞, b/a=2 and $b_h$=0.5$b$.

## 4- Global behavior of the patterns studied numerically

Using the bar-framework numerical approach [4], we perform eigen-value analysis of sample patterns shown in Fig 3 A and B, by changing the ratio of the stiffness of the facet ($K_{facet}$) to that of the fold line ($K_{fold}$). We observe that, similarly to Miura-ori and $BCH_n$ zigzag-base patterns [4, 10], for small values of $K_{facet}/K_{fold}$, which cover a wide range of material properties, the twisting and saddle-shaped modes are the first and the second softest modes, respectively (Fig. 5 A and B). That the second softest bending mode is a saddle-shaped deformation further shows that the material, constructed from a ZCH sheet, has a positive out-of-plane Poisson's ratio (see Fig. 5B and Fig. A1). Moreover, the patterns exhibit rigid origami behavior (Fig. 5D) for large values of $K_{facet}/K_{fold}$ which is in accordance with our expectation.

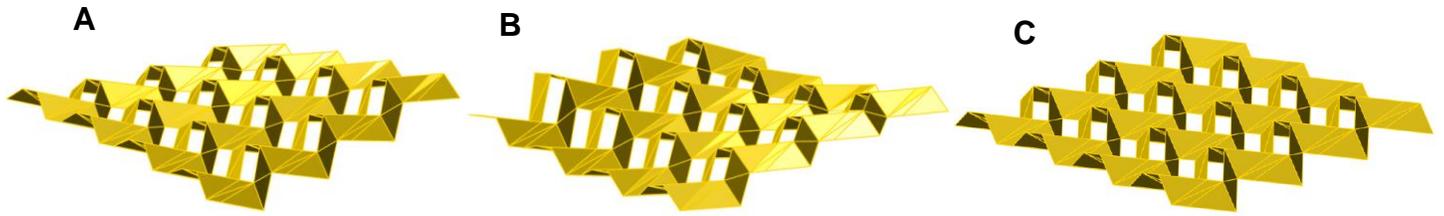

**Fig. A8. Behavior of ZCH sheets under bending and the results of eigenvalue analyses of sample ZCH sheets.** (**A**) Twisting, (**B**) saddle-shaped and (**C**) rigid origami behavior (planar mechanism) of a 4x4 and a 4x3 patterns of ZCH with the holes located on the same directions ($a$=1; $b$=2; $\alpha$=60°).